# High-Mobility Few-Layer Graphene Field Effect Transistors Fabricated on Epitaxial Ferroelectric Gate Oxides

X. Hong[1], A. Posadas[2], K. Zou[1], C. H. Ahn[2] and J. Zhu[1,*]

[1]Department of Physics, The Pennsylvania State University, University Park, PA 16802

[2]Department of Applied Physics, Yale University, New Haven, CT 06520

**Online Supplementary Information Content:**

1. Characterizations of $Pb(Zr_{0.2}Ti_{0.8})O_3$ (PZT) films.
2. Substrate preparation before the exfoliation of graphene.
3. The band structure of FLG.
4. Dielectric constant measurements of PZT.
5. $\rho\,(V_g)$ and $R_H(V_g)$ fitting inside the band overlap regime.
6. The deformation potential of longitudinal acoustic (LA) phonons in graphene.
7. Resistivity and Hall measurements of a $SiO_2$-gated FLG.

## 1. Characterizations of $Pb(Zr_{0.2}Ti_{0.8})O_3$ films

The crystalline quality of the PZT thin films is characterized with x-ray diffraction (XRD). Figure S1 shows the $\theta$-$2\theta$ measurement of a typical 300 nm $Pb(Zr,Ti)O_3$ (PZT) film grown epitaxially on $SrTiO_3$ (STO) substrate. Only 00*n* (*n* =1, 2, etc.) peaks for PZT and STO are detected, showing a *c*-axis oriented film growth, with the polarization *P* pointing normal to the surface (characterized with piezoresponse force microscopy, see below). The *c*-axis lattice constant is ~4.15 Å, consistent with the value of a partially relaxed film. The rocking curve of the 001 peak has a full-width-half-maximum of 0.04°, close to the instrumental resolution, reflecting very high crystallinity.

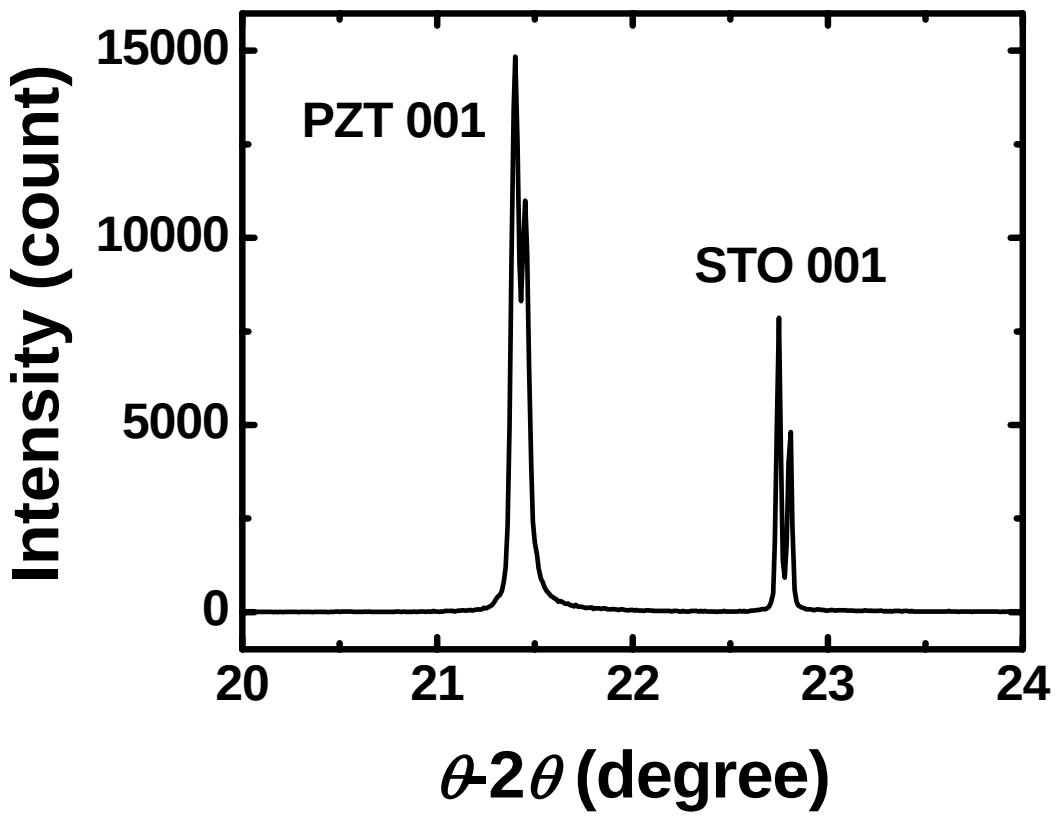

Fig. S1 X-ray $\theta$-$2\theta$ measurement of a 300 nm PZT film grown on a $SrTiO_3$ substrate. The double-peak feature is due to the $k_{\alpha 1}$ and $k_{\alpha 2}$ x-rays.

We have chosen film thickness of $d$ = 300 and 400 nm to enhance the optical visibility of graphene through modeling described in Ref. [1]. AC voltages and voltage pulses of different polarity and magnitude are applied to the AFM tip scanning in contact with the PZT surface to perform piezoresponse force microscopy (PFM) [2]. Negative voltage pulses > 8 V are needed to switch the polarization of 400 nm as-grown films. We conclude that *P* uniformly points into the surface, as expected from the growth procedure. The direction of *P* remained unchanged throughout the current study.

## 2. Substrate preparation before the exfoliation of graphene

Prior to graphene exfoliation, PZT and $SiO_2$ substrates are sonicated in acetone for 20 minutes followed by an IPA rinse (1 min) and drying under a stream of dry $N_2$ gas. They are subsequently baked at 120°C for 5 minutes before the exfoliation of graphene.

### 3. The band structure of FLG

We employ a simply two-band model to estimate the band overlap energy $\delta\varepsilon$ and the densities of electrons and holes in the two-carrier regime near the charge neutrality point of our device, using band parameters determined by Novoselov *et al.* for FLG in Ref. [3] and supporting materials. Measurements there indicate a single electron band with an effective mass $m^*_e = 0.06m_0$ and two hole bands with a heavy hole mass $m^h_h = 0.1m_0$ and a light hole mass $m^l_h = 0.03m_0$. Note these effective mass values are close to those found in bulk graphite: $m^*_e = (0.056\pm0.003)\ m_0$ and $m^*_h = (0.084\pm0.005)\ m_0$ as well [4]. Without knowing the exact offset between the two hole bands, we simplify the estimate by ignoring the light holes, which likely account for less than 25% of the hole population due to the mass ratio. The densities of states of electrons and holes are then given by $\frac{2m^*_{e,h}}{\pi\hbar^2}$, where we have included the 4-fold degeneracy due to valley and spin [3].

The density of electrons and holes in the band overlap regime of the FLG is calculated as follows:

$$n_{e,h} = n^0_{e,h} \pm \frac{\alpha V_g m^*_{e,h}}{(m^*_e + m^*_h)} \qquad \text{(“+” for electron and “–” for hole)} \qquad \text{(S1)},$$

where $n^0_{e,h}$ represents the electron/hole density at the charge neutrality point. At the threshold voltage $V^T_g = 1.1$ V (Fig. 2(b)) where the hole band is completely filled, $n_e = 1.5\text{x}10^{12}/\text{cm}^2$ and $n_h = 0$. Using Eq. S1, we estimate that $n^0_e = n^0_h \sim 9\text{x}10^{11}/\text{cm}^2$ at the charge neutrality point. This translates into a band overlap of $\delta\varepsilon \sim 30$ meV between the electron and hole bands. Alternatively, we determine $\delta\varepsilon$ by fitting $\rho(T)$ at the charge neutrality point to the thermal excitation model described in Ref. [3] and obtain $\delta\varepsilon \sim 27$ meV. The good agreement between these two methods and with theory [5] gave us confidence in these estimates. However we emphasize that the central results of our paper are derived from the electron-only regime ($V_g > V^T_g$) and do not rely on an accurate knowledge of the band structure in the two-carrier regime.

## 4. Dielectric constant measurements of PZT

We deduce the dielectric constant $\kappa$ of PZT substrate by Hall and low-frequency capacitance measurements independently. Figure S2 shows hole densities extracted from Hall measurements as a function of the backgate voltage $V_g$ in three different devices placed on the same PZT substrate, including the device shown in Fig. 2. We determine the charge injection rate $\alpha = (1.35\pm0.05)\text{x}10^{12}$ cm$^{-2}$/$V_g$(V) using Eq. (2) in the single carrier (hole) regime, and calculate $\kappa \approx 100$ using a parallel-plate capacitor model.

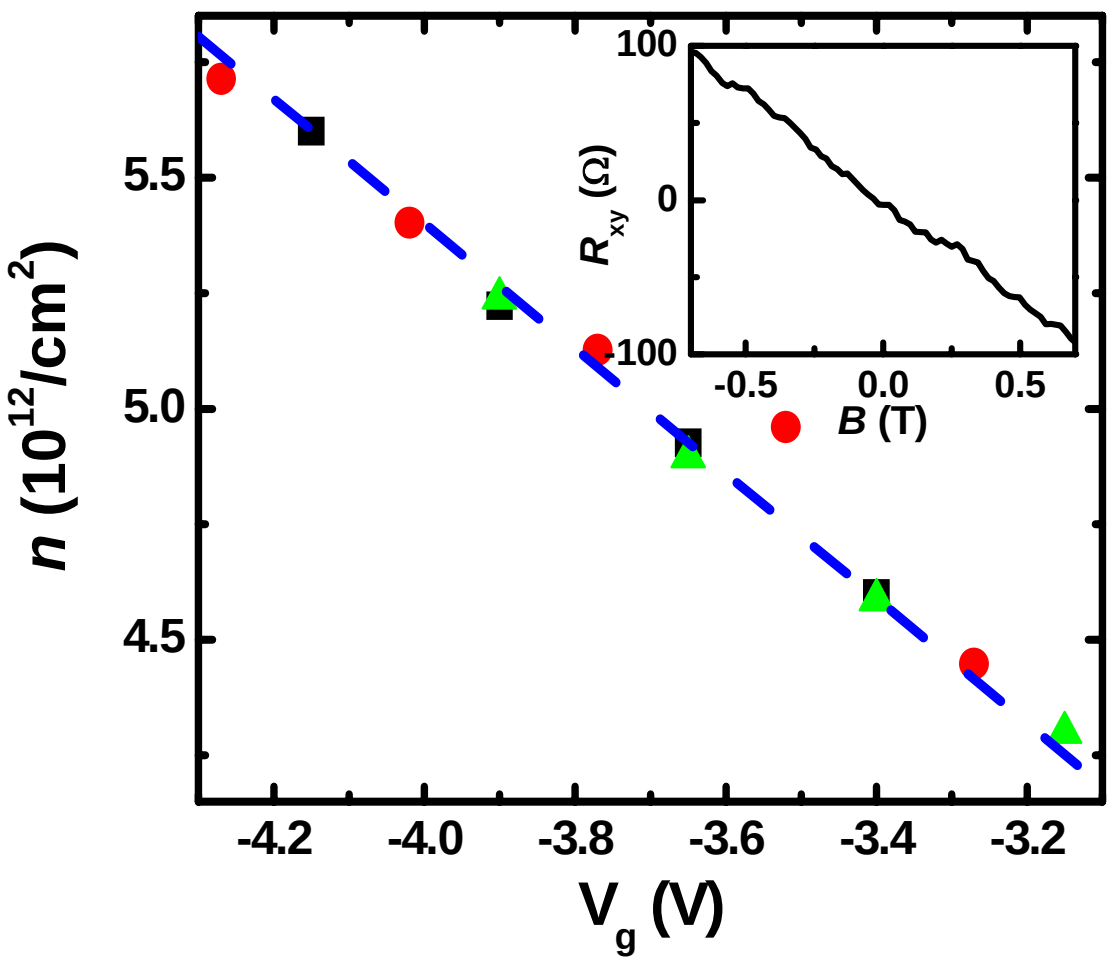

Fig. S2 Hole density $n$ vs. backgate voltage $V_g$ taken on three devices (squares, circles and triangles) on the same PZT substrate. The fit (blue dashed line) corresponds to a charge injection rate of $1.35\text{x}10^{12}$ cm$^{-2}$/V. $V_g$ is shifted to align the charge neutrality point at $V_g = 0$ V as described in the text. Inset: Hall resistance $R_{xy}$ vs. perpendicular magnetic field for $n = 4.6\text{x}10^{12}/\text{cm}^2$.

We also determine $\kappa$ directly through low-frequency (20-1000 Hz) capacitance measurements. Capacitors with varying areas are measured to eliminate the effect of parasitic capacitances. We calculate the area of a capacitor from the mask design and estimate $\kappa \approx 120$ from these measurements.

**5. $\rho$ ($V_g$) fitting inside the band overlap regime**

We fit our low-*T* $\rho$ ($V_g$) data to Eqs. (1) and (3) within the band overlap regime, using effective masses of $m^*_e = 0.06m_0$ and $m^*_h = 0.1m_0$. A different set of mass values may lead to a different band overlap $\delta\varepsilon$ and affects the value of the exponent $\beta$ in Eq. (3). For example, $\delta\varepsilon \sim 32$ meV and $\beta \sim 1.0$, or $\delta\varepsilon \sim 27$ meV and $\beta \sim 0.8$ can fit the same data equally well as the values used in the text ($\delta\varepsilon \sim 30$ meV and $\beta \sim 0.9$). Beyond this energy range of $\delta\varepsilon$ (27-32 meV), a power-law *n*-dependence of mobility no longer describes the data well. The above analysis shows that although the precise value of $\beta$ depends on $\delta\varepsilon$ and hence the detailed understanding of the band structure of FLG, the power-law *n*-dependence seems to be robust.

This *n*-dependence of mobility is also supported by the measurements of the Hall coefficient $R_H$. Within the band overlap regime, $R_H$ exhibits the characteristics of a two-carrier system described by Eq. S2:

$$R_H = \frac{n_h\mu_h^2 - n_e\mu_e^2}{e(n_h\mu_h + n_e\mu_e)^2} \qquad \text{(S2).}$$

At high $|V_g|$, the system becomes a purely electron or hole gas and $R_H$ reads:

$$R_H = \frac{1}{en_{e,h}} = \frac{1}{e\alpha(V_g - V_g^0)} \qquad \text{(S3),}$$

Where for PZT-gated samples $\alpha = 1.35\text{x}10^{12}$ cm$^{-2}$/V and $V^0_g$ is offset to 0 V. Figure S3 plots $R_H$ vs. $V_g$ of the FLG shown in Fig. 2, together with three fitting curves. $R_H$ calculated using Eq. (S2) with $n_{e,h}$ given by Eq. (S1) and $\mu_{e,h}$ given by Eq. (3) with $\beta = 0.9$ and $r = 0.6$ (red) produce an excellent agreement with the data. Calculations based on a constant mobility $\mu_h = \mu_e$ = const. (green), or an $n$-dependent but symmetric electron and hole mobility $\mu_h = \mu_e \sim n^{0.9}$ (blue) clearly do not fit the data.

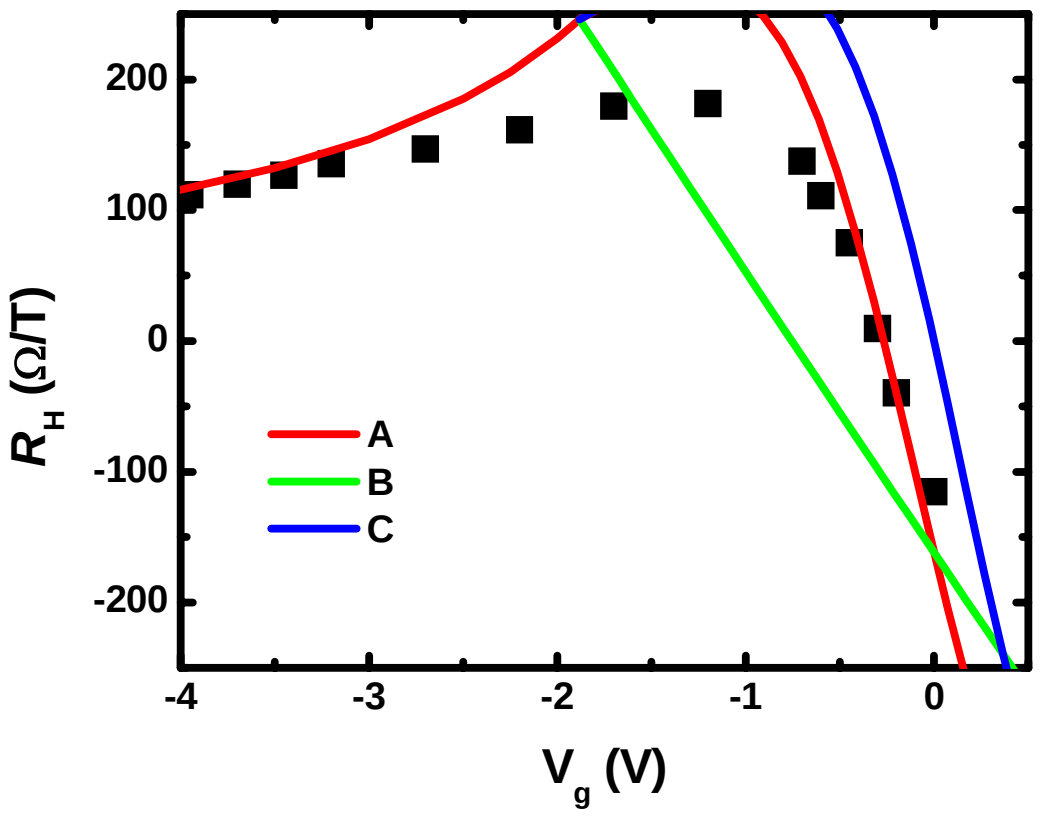

Fig. S3 Hall coefficient $R_H$ vs. $V_g$ at 10 K with three calculated curves. Curve A (red): $\mu_{e,h} \sim n_{e,h}^{0.9}$, $\mu_h/\mu_e = 0.6$ (Eq. 3 of text). Curve B (green): $\mu_h = \mu_e$ = const. Curve C (blue): $\mu_h = \mu_e \sim n^{0.9}$. The charge injection rate $\alpha = 1.35\text{x}10^{12}\text{cm}^{-2}/\text{V}$.

## 6. The deformation potential of LA phonons in graphene

In Eq. (4), we assume an unscreened deformation potential $D$ for LA phonons. Studies in GaAs show that neglecting the dielectric screening may result in an underestimate of the deformation potential [6]. The detailed $T$-dependence in the BG regime from a sample of yet higher mobility is required to determine the effect of screening. This analysis is beyond the scope of the present paper.

## 7. Resistivity and Hall measurements of a $SiO_2$-gated FLG

To compare the performance of PZT and $SiO_2$ substrates, we fabricate a FLG-FET of the same thickness (2.4 nm) on 300 nm $SiO_2$ following identical preparation steps. Figure S4 shows $\rho(V_g)$ of the $SiO_2$-gated FLG at selected temperatures. The charge neutrality point occurs at $V^0_g$ = - 15.5 V due to unintentional chemical doping. We find the charge injection rate of the backgate to be $\alpha = 7.0\text{x}10^{10}$ $\text{cm}^{-2}/\text{V}$, in excellent agreement with calculations and experimental values from other graphene FETs made from the same set of wafers. We determine $\mu(T)$ using Eq. (2) at electron density $n = 2.4\text{x}10^{12}$ cm$^{-2}$ ($V_g$ = 19 V). The results are plotted in Fig. 4 as open triangles. The mobility $\mu$ is approximately 14,000 cm$^2$/Vs and shows a very weak $T$-dependence. These findings are in good agreement with what has been reported in the literature [3].

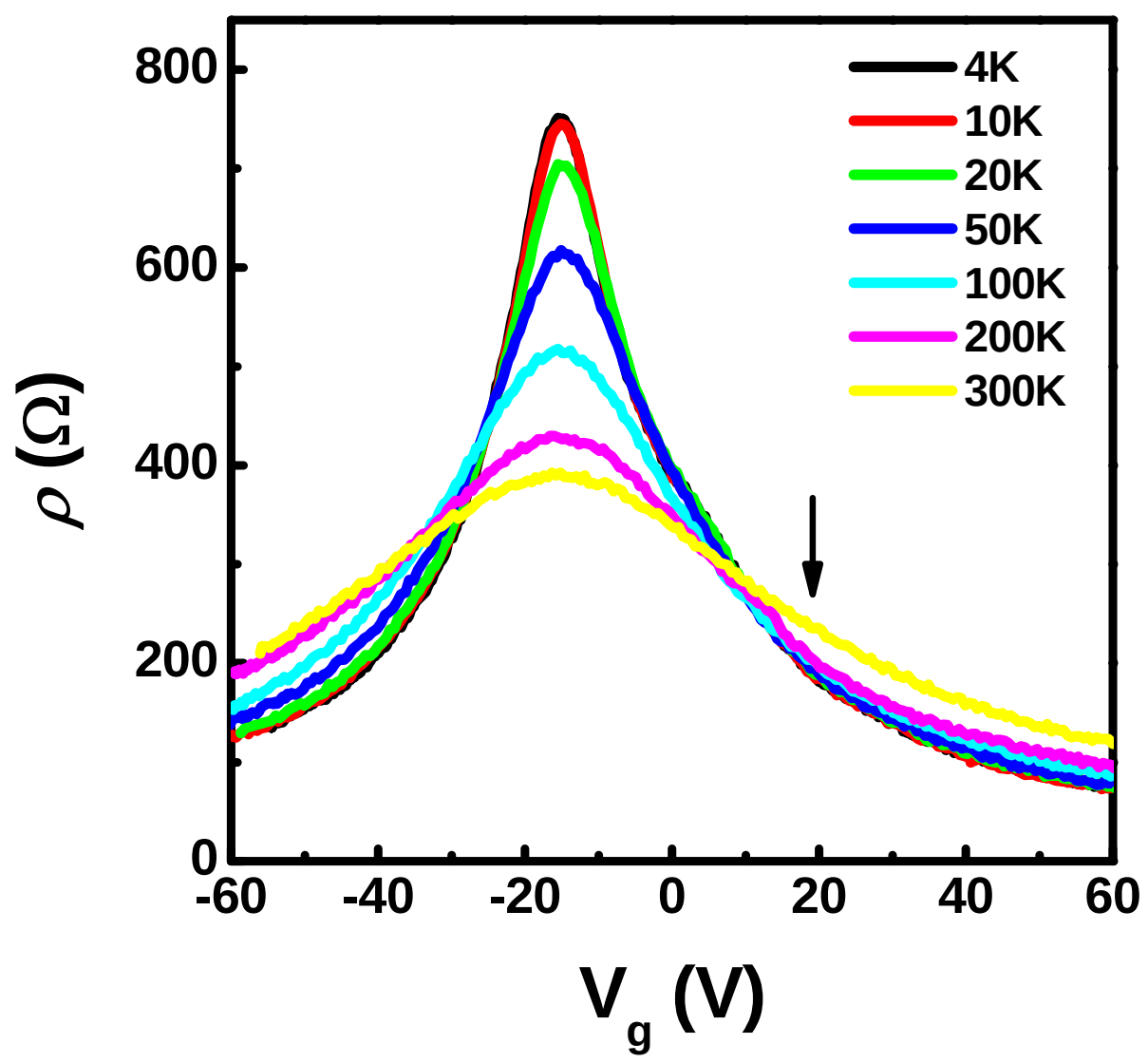

Fig. S4 $\rho(V_g)$ of the $SiO_2$-gated FLG (2.4 nm) at selected temperatures 4 K < $T$ < 300 K. The charge neutrality point $V^0_g$ = - 15.5 V. The arrow marks $V_g$ = 19 V corresponding to $n = 2.4\text{x}10^{12}$ cm$^{-2}$, where we determine the mobility plotted in Fig. 4.

Figure S5 shows $R_H$ vs. $V_g$ in this FLG. The band overlap, electron-only and hole-only regimes are all clearly visible. Red solid lines are calculated from Eq. (S3) using $\alpha = 7.0\text{x}10^{10}$ cm$^{-2}$/V and $V^0_g$ = - 15.5 V. The transition voltage to the electron-only regime (~ 2 V) is in good agreement with that of the PZT-gated FLG, indicated by an arrow.

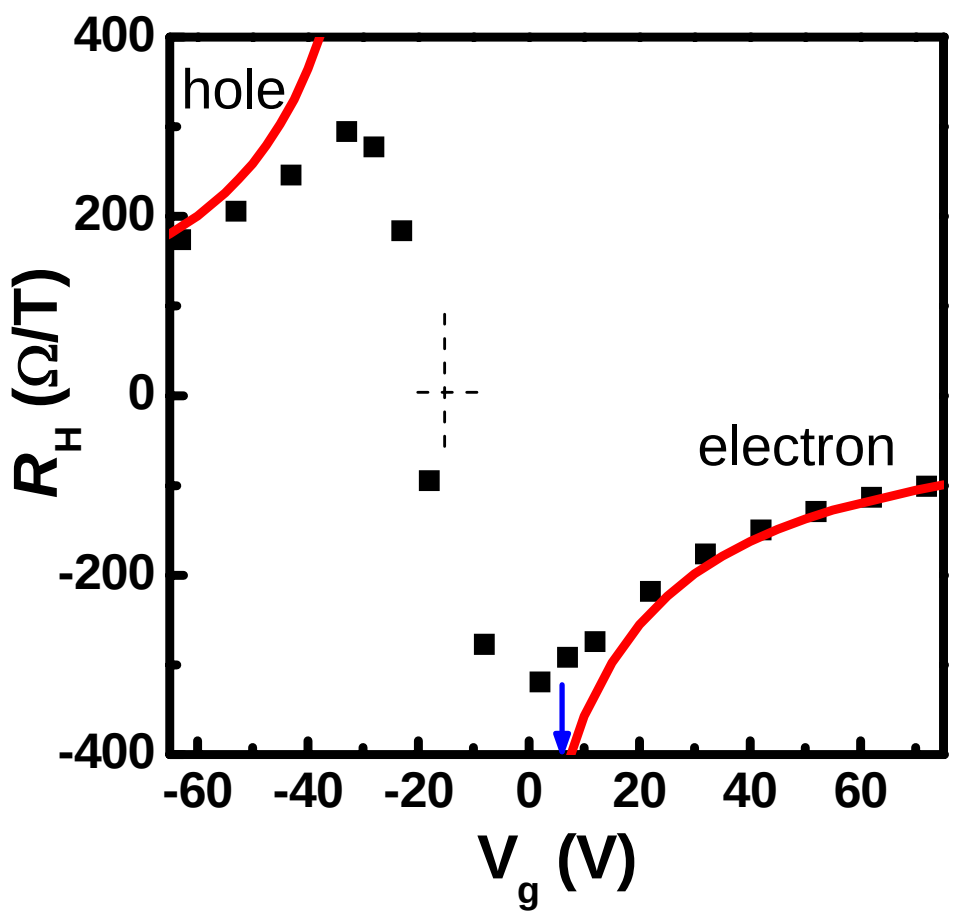

Fig. S5 Hall coefficient $R_H$ vs. $V_g$ ($T$ = 4 K) of the $SiO_2$-gated FLG. Red solid lines are calculated using Eq. S3 and correspond to electron- and hole-only regime, respectively. The blue arrow indicates the equivalent voltage of the band edge observed in the PZT-gated FLG of the same thickness. The cross marks (-15.5V, 0 Ω/T).